\documentclass[aps,reprint,superscriptaddress,nofootinbib,showpacs]{revtex4-1}
\usepackage{amsmath,latexsym,amssymb,hyperref,graphicx}

\hypersetup{colorlinks,citecolor= nicegreen,linkcolor= nicered}
\usepackage{color}
\usepackage{hyperref}
\definecolor{red}{rgb}{1,0,0}
\definecolor{nicered}{rgb}{0.7,0.1,0.1}
\definecolor{nicegreen}{rgb}{0.1,0.5,0.1}

\newcommand\GeV{\text{GeV}}
\newcommand\TeV{\text{TeV}}

\newcommand\pb{\text{pb}}
\newcommand\fb{\text{fb}}

\newcommand\SEC[1]{\medskip\noindent{\sl\bfseries #1}}

\begin{document}

\title{First Limits on Left-Right Symmetry Scale from LHC Data}

\author{Miha Nemev\v{s}ek}
\affiliation{ICTP, Trieste, Italy}
\affiliation{Jo\v zef Stefan Institute, Ljubljana, Slovenia}

\author{Fabrizio Nesti}
\affiliation{Universit\`a di Ferrara, Ferrara, Italy}

\author{Goran Senjanovi\'{c}}
\affiliation{ICTP, Trieste, Italy}

\author{Yue Zhang}
\affiliation{ICTP, Trieste, Italy}

\date{\today}

\vspace{1cm}

\begin{abstract}
\noindent
We use the early Large Hadron Collider data to set the lower limit on
the scale of Left-Right symmetry, by searching for the right-handed
charged gauge boson $W_R$ via the final state with two leptons and
two jets, for $33\,\pb^{-1}$ integrated luminosity and $7\,\TeV$
center-of-mass energy. In the absence of a signal beyond the Standard
Model background, we set the bound $M_{W_R} \gtrsim 1.4\,\TeV$ at 95\%
C.L.. This result is obtained for a range of right-handed neutrino
masses of the order of few 100\,GeV, assuming no accidental
cancelation in right-handed lepton mixings.
\end{abstract}



\maketitle

\SEC{Introduction.} For more than three decades, the left-right (LR) symmetric gauge theories~\cite{lr}
have been one of the most popular extensions of the Standard Model (SM), introduced originally for
the sake of understanding the breakdown of parity in weak interactions. These theories played a
profound role in the development of neutrino mass. They required non-vanishing neutrino mass long
before it was to be confirmed experimentally and moreover, they led to the
seesaw mechanism~\cite{minkowskims, seesaw}, nowadays a well-established framework of small neutrino
mass. 

The question is at which scale does the LR symmetry get restored, or
equivalently, what is the mass of the right-handed charged gauge boson
$W_R$.
%
In the minimal model, there exist strong theoretical limits~\cite{Beall:1981ze} 
on the scale of the theory from $K_L-K_S$ mass difference . This limit depends on the choice between two
possible discrete left-right symmetries: parity ($\mathcal P$) and
charge conjugation ($\mathcal C$). In the case of $\mathcal P$, the
limit is $M_{W_R} \gtrsim 3\,\TeV$, whereas in the case of $\mathcal C$
it is somewhat lower: $m_{W_R} \gtrsim 2.5\,\TeV$~\cite{Maiezza:2010ic}.


The seesaw version of the theory offers a particularly exciting signal
in the form of the lepton-number breaking channel of a same-sign
lepton pair and two jets without missing energy~\cite{Keung},
intimately related with the Majorana nature of neutrino
mass. Dedicated studies of both, ATLAS and CMS, show that the LHC
running at 14\,\TeV\ can reach $M_{W_R} \lesssim 2(4)\,\TeV$ with a
luminosity of $0.1(30)\,\fb^{-1}$~\cite{Ferrari,Gninenko:2006br}.


Moreover, the LR scale may well be required to lie in the Large Hadron
Collider (LHC) energy region. The point has to do with the
neutrino-less double beta decay, which has been claimed to have been
observed~\cite{KlapdorKleingrothaus:2004wj}.  One possible source of
this process is a Majorana neutrino mass, but if cosmology keeps
pushing down the sum of neutrino masses and if this claim is to be
confirmed, new physics behind neutrino-less double beta decay would be
a must.  LR symmetry plays naturally that
role~\cite{Mohapatra:1980yp}, and this would require the LR scale to
be in the TeV region. One could have a profound interplay between high
energy collider experiments and low energy neutrino-less double beta
decay~\cite{Tello:2010am}.

In spite of the short period of running and a fairly low luminosity,
the sensitivity achieved by both ATLAS and CMS collaborations, allows
one to already set relevant updated bounds on a number of new
particles and their interactions.
%
For example, in case the right-handed neutrinos are very light, or
equivalently, neutrinos being the Dirac particles, the right-handed
charged boson decays leptonically in the manner often associated with
the nomenclature $W' \to \ell \nu$. Recently, the CMS collaboration
established the generic bound for such particles $M_{W'} \gtrsim
1.4\,\TeV$, for the same couplings of $W$ and
$W'$~\cite{Khachatryan:2010fa, Collaboration:2011dx}.

%
Inspired by this, we investigate carefully the analogous limit in the
Majorana case of the LR theory, by using the available LHC data. This
theoretically preferred scenario, which requires heavy right handed
neutrinos, leads at this stage to a very similar bound, $M_{W_R}
\gtrsim 1.4\,\TeV$ at 95\% confidence level (CL) for a large
portion of parameter space.  In particular, this applies to
right-handed (RH) heavy neutrinos in the LHC accessible region, $m_N =
\mathcal O (100)\,\GeV$ and generic RH lepton flavor mixing
angles.


This lower bound would go to 2.2\,TeV for $\sqrt s = 7\,\TeV$,
and a luminosity of $1\,\fb^{-1}$.

\SEC{The generic gauge structure.} The minimal LR symmetric theory is
based on the gauge group $\mathcal G_{LR} = SU(2)_L \times SU(2)_R
\times U(1)_{B-L}$ (suppressing color), with corresponding gauge
couplings $g_L$, $g_R$ and $g_X$, and a symmetry between the left and
right sectors.  Quarks and leptons come in LR symmetric
representations $q_{L,R }= (u, d)_{L,R}$ and $\ell_{L,R} = \left( \nu,
  e \right)_{L,R}$.

At this point, it is sufficient to assume that this gauge symmetry is
broken down to the SM at a scale $M_R$.  If $g_L \approx g_R$, the
Tevatron sets a rough bound (to be discussed below more carefully)
$M_{R}\gtrsim\TeV$.  This is enough to ensure a small mixing angle
between left and right gauge bosons, which for all practical purposes
is taken to be zero in what follows.

\begin{figure*}[t!]
  \centerline{\includegraphics[width=1.8\columnwidth]{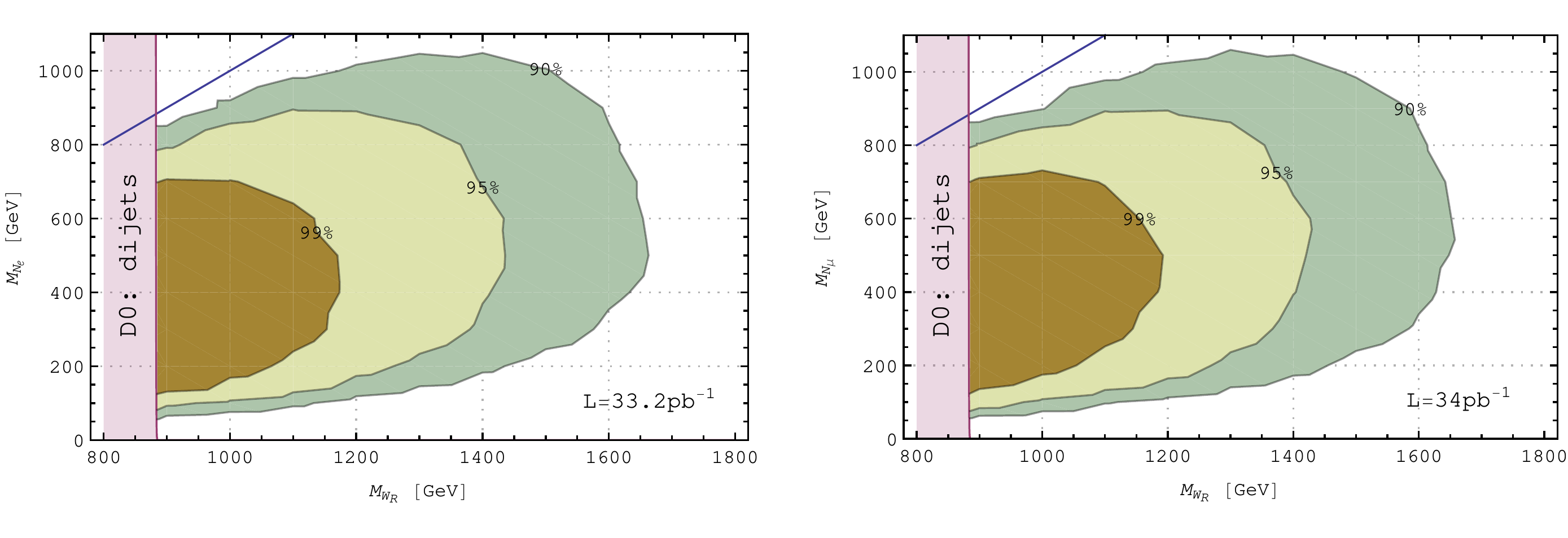}}\vspace{-4ex}%
  \caption{Exclusion (90\%, 95\%, 99\% CL) in the $M_{W_R}$--$m_N$
    plane from the $ee jj$ (left) and $\mu\mu jj$ (right) channel. We
    assume no accidental cancelation in the RH lepton mixings. The
    2$\sigma$ lower bound $\sim$1.4\,TeV is valid over a range of RH
    neutrino masses of order several hundred GeV.}
  \label{figLimitsMWRmN}
\end{figure*}

The physical gauge fields consist of the usual the SM states and the
new ones: $W_R^\pm$ and $Z_{LR}$, with the following interactions
\begin{eqnarray}
\frac{g_R}{\sqrt2} W_{R\mu}^+ \left[ \bar u_R \gamma^\mu d_R + \bar\nu_R \gamma^\mu \ell_R \right] + {\rm h.c.} \ ,
\end{eqnarray}
where we suppress the family indices, together with the flavor mixing indices, and 
%
%
$ \frac{g_R}{\sqrt{1 - \tan^2 \theta_W g_L^2 / g_R^2}}Z_{LR}^{\mu}
\bar f \gamma_\mu \left[ T_{3R} + \tan^2 \theta_W \frac{g_L^2}{g_R^2}
  (T_{3L} - Q) \right] f$,
where $\theta_W$ is the usual weak mixing angle. It is easy to show
that there is a lower limit on $g_R>g_L\tan\theta_W$.  All of this is
independent of the choice of the Higgs sector, responsible for the symmetry
breaking.
What does depend on the choice of the Higgs sector,
is the ratio of $Z_{LR}$ and $W_R$ masses, just as in the SM.

\medskip

Before delving into the Higgs swamp, let us discuss the generic limits
on the new gauge boson masses, most of which depend crucially on the
nature of the right-handed neutrinos.
%
There is one limit on the mass of $W_R$ which depends only on the
value of $g_R$ and the right-handed quark mixing, from the $W_R \to t
b$ channel. Tevatron gives this bound for the same left and right
parameters: $M_{W_R} \gtrsim 885\,\GeV$~\cite{Abazov:2011xs}.

The limit on $Z_{LR}$ mass depends only on $g_R$ and for equal left
and right couplings, and the present limit set by $Z_{LR} \to \mu^+
\mu^-$ and $ee$ channel: $M_{Z_{LR}} \gtrsim 1050\,\GeV$~\cite{CMSCollaboration:2011wq}.

\medskip

%
%
\SEC{The Majorana connection.}  We start first with the seesaw
scenario in which the right-handed neutrinos are heavy Majorana
particles that we denote $N$ in what follows. In a reasonable regime
$10\,\GeV\lesssim m_N \lesssim M_{W_R}$, this opens an exciting lepton
number violating channel~\cite{Keung} $W_R \to \ell^\pm \ell^\pm j j$,
which allows one to probe higher values of
$M_{W_R}$~\cite{Ferrari,Gninenko:2006br}. After being produced through
the usual Drell-Yan process, $W_R$ decays into a charged lepton and a
right-handed neutrino $N$. Since $N$ is a Majorana particle, it decays
equally often into another charged lepton or anti-lepton, together
with two jets.  Ideally, one would like to study both same sign lepton
pairs, for the sake of lepton number violation, and any sign lepton
pair for the sake of increasing the sensitivity of the $W_R$ search.

Such a final state with any-sign lepton pair was used recently by the
CMS collaboration to search for pair production of scalar leptoquarks,
for both electron and muon lepton flavors~\cite{Khachatryan:2010mp}.
We thus use these data to impose an improved limit on the
masses of $W_R$ and $N$~\cite{lq}.

We perform a Monte Carlo simulation, using {\sc MadGraph~\cite{Alwall:2007st}, \sc Pythia}~\cite{Sjostrand:2007gs}
to generate the events for the process $ p p \to n \ell \, n' j \, (n, n' \geq
2)$
%
%
and do the showering, including the K-factor of 1.3 to account for the
NNLO QCD corrections~\cite{Hamberg:1990np}. We simulate the CMS
detector using both {\sc PGS} and {\sc Delphes} which give essentially
the same result. We also use the CTEQ6L1 parton distribution functions
(PDF).  We summarize in Table~\ref{cuts} the cuts used in this Letter,
taken from the CMS papers~\cite{Khachatryan:2010mp}.  For jet
clustering, we employ the {\sc FastJet}
package~\cite{Cacciari:2005hq}, using the anti-$k_T$ algorithm with
$R=0.5$ for jet reconstruction.  The lepton isolation cut makes our
exclusion less efficient in the region of light $N$, roughly below
50\,GeV. The reason is that the lepton and jets coming from the
boosted $N$ decay become too collimated and finally merge into a
single jet with a lepton inside. However, when $N$ is heavier, this
cut becomes less relevant.

\begin{table}[b]
\begin{tabular}{|c|c|c|c|c|c|}
\hline
\ channel \        &  $p_T^{\min}(\ell)$ &  $|\eta(\ell)|^{\max}$ & $\Delta R(\ell, \ell)^{\min}$ & $m^{inv}_{\ell \ell}$&$S_T$ \\
\hline
$e e j j$         &  $ 30\,\GeV$\  & 2.5 & 0.3 &  $125\,\GeV$\ &\ optimal\ \ \\
$\mu \mu jj$  &  $ 30\,\GeV$\  & 2.4 &  0.3 &  $115\,\GeV$\ &\  optimal\ \ \\
\hline
\end{tabular}
\caption{In both cases we also demand at least two jets with $p_T > 30\,GeV$
and $|\eta| < 3$. Moreover, in the $\mu \mu jj$ case, at least one muon has to be within $|\eta|<2.1$ and
in the $ee j j$ case both electrons have to be separated from either jet by $\Delta R (e,j) > 0.7$.}
\label{cuts}
\end{table}

The data and the SM background are taken from
Ref.~\cite{Khachatryan:2010mp}. The main contributions to the
background include the $t \bar t + \text{jets}$ and $Z/\gamma^* +
\text{jets}$ and they can be suppressed efficiently by the appropriate
cut on the invariant mass of the two leptons ($m_{ee} > 125\,\GeV$ and
$m_{\mu \mu} > 115\,\GeV$).
%
%
We employ the Poisson statistics to get the exclusion plots.  In order
to get the most stringent bound, for each point in the $M_{W_R}-m_N$
parameter space we choose the optimal cut on the $S_T$ parameter (the
scalar sum of the $p_T$ of the two hardest leptons and the two hardest
jets) from Table 1 of~\cite{Khachatryan:2010mp}.

The resulting $95\%\,\text{CL}$ limit $M_{W_R} \gtrsim 1.4\,\TeV$, the
best up to date, holds for a large portion of parameter space, as
shown in Fig.~\ref{figLimitsMWRmN}. One can see that this result
restricts the RH neutrino masses to lie roughly in a fairly natural
energy scale 100\,GeV--1\,TeV.  It turns out that both the electron
and muon flavour channels give a similar exclusion in the parameter
space.

In all honesty, this limit could be weakened by a judicial choice of RH
leptonic mixing angles and phases; we opted here against such
conspiracy. For example, in the case of an appealing type-II seesaw,
left and right leptonic mixing angles are related to each other, and
no suppression arises~\cite{Tello:2010am}.  A careful study of the
mixings, through {\it e.g.} flavor-changing $e\mu$ final
state~\cite{CMSCollaboration:2011wq} 
will be published elsewhere.

\begin{figure}[t!]
  \centerline{\includegraphics[width=.7\columnwidth]{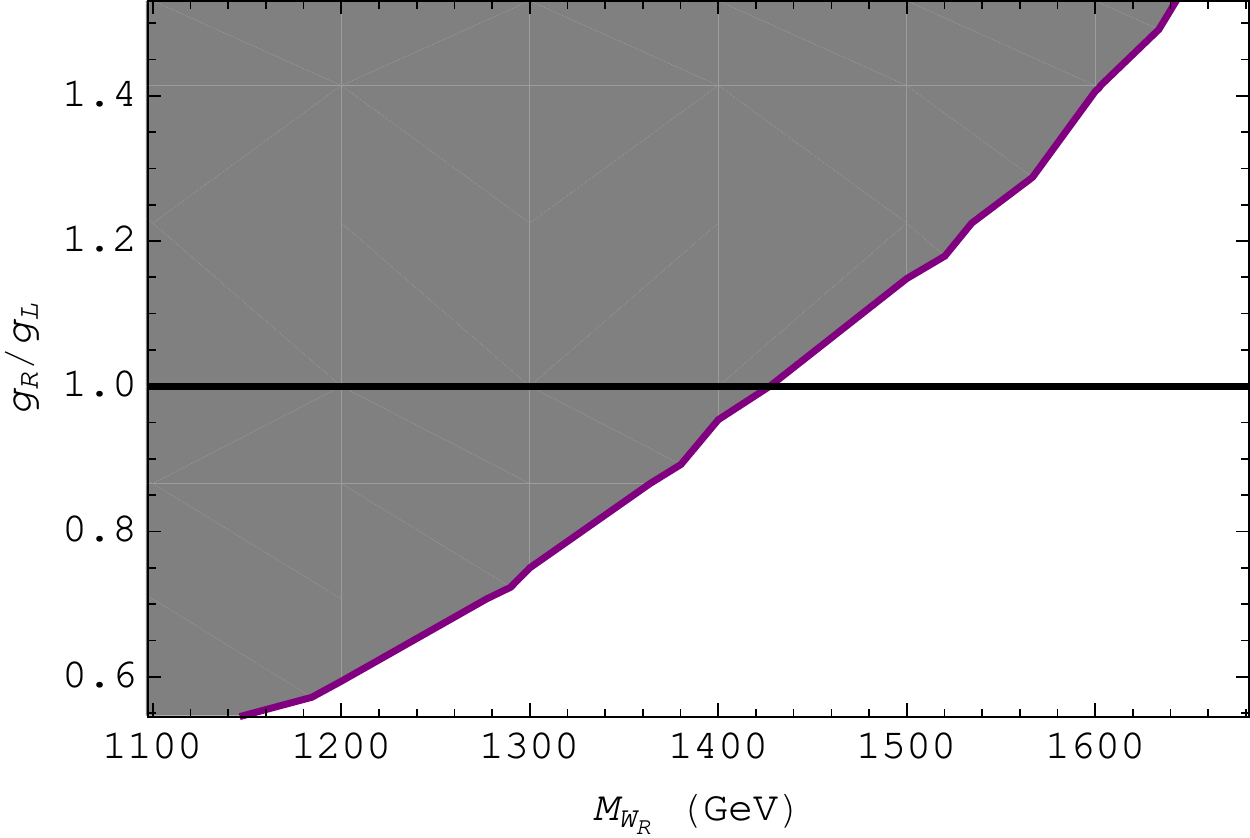}}\vspace*{-2ex}
  \caption{Here we vary the ratio $g_R/g_L$. The shaded region is the 95 \% CL exclusion on $W_R$ mass
  for fixed value of the RH neutrino mass, chosen illustratively to be  $m_N=500\,\GeV$.}
  \label{gRgL}
\end{figure} 
 
Up to now, we have made an assumption that $g_R=g_L$ and the right-handed counterpart of the
Cabibbo angle is the same as the left-handed one. This is actually true in the minimal version of the LR symmetric 
theory, but need not be so in general. One could easily vary the right-handed quark mixing parameters, but 
the presentation would become basically impossible with so many parameters and different PDF sets.
We relax though the $g_R=g_L$ assumption since this captures the essence of the impact when right and 
left are different. In the Fig.~\ref{gRgL}, in the shaded area, we plot the 95 \% CL exclusion region in the
$g_R / g_L$ versus $M_{W_R}$ plane, for a fixed value $m_N= 500\,\GeV$. Clearly, with the increased
$g_R$ the production rate goes up and so does the limit on the mass of the right-handed gauge boson.

%
%
\SEC{The Dirac connection.}  
In case the right-handed neutrinos are very light, they are treated as missing energy at the LHC and this case 
is equivalent to the case of Dirac neutrinos to which we now address our attention.
 This is actually the original version~\cite{lr} of the LR symmetric
theory, not popular anymore precisely since the neutrinos end up being Dirac particles. In this case
the best limit comes from the recent CMS studies of $W' \to e \nu$ decay~\cite{Khachatryan:2010fa}:
$M_{W_R} \gtrsim 1.36\,\TeV$ and $W'\to \mu\nu$ decay~\cite{Collaboration:2011dx}: 
$M_{W_R} \gtrsim 1.4\,\TeV$. Even with a low luminosity, LHC is already producing
a better limit than the Tevatron one: 1.12\,TeV~\cite{Aaltonen:2010jj}. 
%
  
 \SEC{The Higgs connection.}   We discuss briefly the minimal models of Majorana and Dirac cases.

{\em Majorana neutrino}.  
The Higgs sector\footnote{There is also a bidoublet, which takes the
  role of the SM Higgs doublet, and we do not discuss it here.  For a recent
   discussion of the limits on its spectrum and 
   phenomenology, see~\cite{Maiezza:2010ic}.} consists of~\cite{minkowskims}: the $SU(2)_{L,R}$
triplets  $\Delta_L $ and $\Delta_R $.
Besides giving a Majorana mass to $N$, a non-vanishing $\langle \Delta_R \rangle$ 
leads to the relation between the new neutral and charged gauge bosons
\begin{eqnarray}
\frac{M_{Z_{LR}}}{M_{W_R}} = \frac{\sqrt2 g_R/g_L}{\sqrt{ (g_R / g_L)^2 - \tan^2\theta_W }} \ .
\end{eqnarray}
For $g_R\approx g_L$, one gets ${M_{Z_{LR}}}\approx 1.7 \, {M_{W_R}}$. In
this case, one can infer the lower bound on $M_{Z_{LR}}$ from the lower bound on $M_{W_R}$
in Fig.~\ref{figLimitsMWRmN} and it exceeds the direct search result 
from~\cite{CMSCollaboration:2011wq}. For example, in the case of $m_N \sim 500\,\GeV$,
the $Z_{LR}$ with a mass below 2.38\,TeV is excluded.

%
%
 
{\em Dirac neutrino}. 
In this case, the triplets are traded for the
usual SM type left and right doublets, as in the original version of
the LR theory~\cite{lr}. For us the only relevant change is the ratio
of heavy neutral and charged gauge boson masses, which goes down by a
$\sqrt2$.

%

\begin{figure*}[t!]
  \centerline{\includegraphics[width=2\columnwidth]{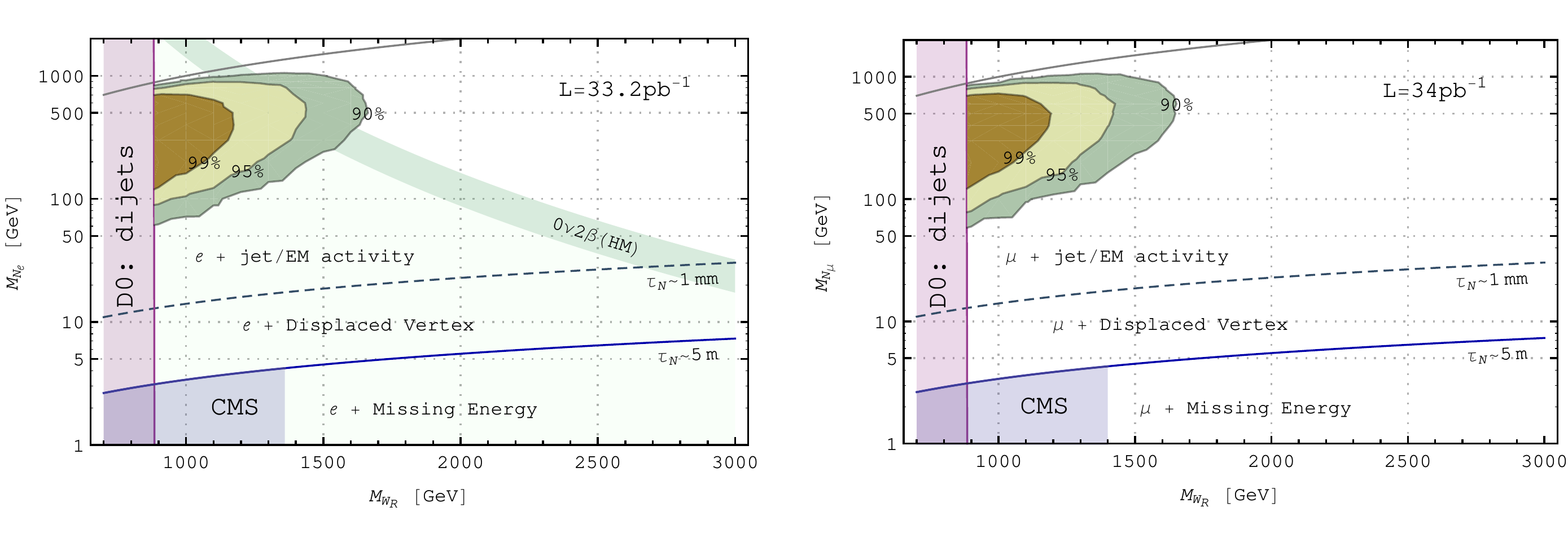}~~}\vspace{-3.5ex}
  \caption{Limits in the $M_{W_R}$--$m_{N}$ parameter space. The
    elliptical regions correspond to Fig.~\ref{figLimitsMWRmN}, now
    shown in logarithmic scale. The excluded vertical region on the
    left is the D0 result from $W_R \to t b$ decay. The excluded lower
    trapezoid is the CMS missing energy result applicable for neglibly
    small $m_N$.  For illustration in the left plot we also depict a
    (green) band where, ignoring leptonic mixings, the LR contribution
    to $0\nu\beta\beta$ decay saturates the HM
    claim~\cite{KlapdorKleingrothaus:2004wj}. \vspace*{-2ex}}
  \label{figCombLimits}
\end{figure*}

\SEC{Improved limits from CMS data.}  The constraints from the recent
CMS data are shown in Fig.~\ref{figCombLimits}, where the missing
portion of the parameter space, not yet excluded by present data, is
clearly seen. We use the {\sc BRIDGE}~\cite{Meade:2007js} with {\sc
  MadGraph} to calculate the average decay length of (boosted) $N$ in
the low mass region.

We find that for $m_N \lesssim 3-5\,\GeV$, the average decay length
exceeds the size of the detector and is therefore regarded as missing
energy.

The region above it, until about $m_N \lesssim 10-15\,\GeV$,
corresponds to the displaced vertex regime and it has clear signatures
for future discovery.

The white region further above unfortunately still requires published
data or a dedicated analysis in order to set a bound on the $W_R$
mass. This missing region can be easily filled with the data on the
single lepton plus jet with electromagnetic activities (or a muon
inside)~\cite{Ferrari}.

\SEC{Summary and outlook.}  The direct limits on the scale of LR
symmetry up to now have been much below the theoretical limit $M_{W_R}
\gtrsim 2.5\,\TeV$~\cite{Maiezza:2010ic}, but with the advent of
the LHC it is a question of (short) time that the experiment finally
does better.

Moreover, as discussed recently in~\cite{Tello:2010am}, there is an
exciting connection between the high energy collider and low energy
experiments, with the LR scale possibly at the LHC reach.  Motivated
by this, we have used the existing CMS data to set a correlated limit
on the mass of the right-handed charged gauge bosons and right-handed
neutrinos. For reasonable values of right-handed neutrino masses, one
gets $M_{W_R} \gtrsim 1.4\,\TeV$ at 95\% CL and 1.7\,TeV at 90\% CL.

This is comparable to the recent CMS bound $M_{W_R} \gtrsim
1.36\,(1.4)\,\TeV$, applicable to Dirac neutrinos (and/or small
Majorana RH neutrino masses). It is reassuring that the limit seems
quite independent of the nature of neutrino mass. As the luminosity
increases up to $\mathcal L= 1\,\fb^{-1}$, expected by the end of this
year, one could push the limit on $M_{W_R}$ all the way up to 2.2 TeV.
Also, using the data on the dilepton resonance
search~\cite{CMSCollaboration:2011wq} (ignoring the jets), one can set
a limit on $W_R$ in a similar way as discussed in the present work.

There is a window for high values of $m_N$, when all of the right-handed neutrinos are much heavier
than $W_R$. This less likely possibility, if true, will be covered by the limit from the future di-jet (or $t b$) data.




\SEC{Acknowledgments.}  We wish to acknowledge H. An and S.L. Chen for
discussions and help with the {\sc PGS 4}.  F.N. thanks ICTP,
Y.Z. thanks INPAC, Shanghai Jiao Tong University, and F.N., M.N.,
Y.Z. thank BIAS, Trieste for hospitality where most of this work was
performed. We are grateful to Alejandra Melfo and Ivica Puljak for
their interest and encouragement.


\begin{thebibliography}{99}

\bibitem{lr}
J.C.~Pati, A.~Salam,
Phys.\ Rev.\ D {\bf 10} (1974) 275;
R.M.~Mohapatra, J.C.~Pati,
Phys.\ Rev.\ D {\bf 11} (1975) 2558;
G.~Senjanovi\'{c}, R.N.~Mohapatra,
Phys.\ Rev.\ D {\bf 12} (1975) 1502.
G.~Senjanovi\'c,
Nucl.\ Phys.\ B {\bf 153} (1979) 334.


\bibitem{minkowskims}
P.~Minkowski,
Phys.\ Lett.\ B {\bf 67} (1977) 421;
R.N.~Mohapatra, G.~Senjanovi\'{c},
Phys.\ Rev.\ Lett. {\bf 44} (1980) 912.

\bibitem{seesaw}
T.~Yanagida, {\em Workshop on unified theories
and baryon number in the universe}, ed.
A. Sawada, A. Sugamoto (KEK, Tsukuba, 1979);
S.~Glashow, {\em Quarks and leptons,  Carg\`ese 1979}, 
ed. M. L\'evy (Plenum, NY, 1980);
M.~Gell-Mann \emph{et al.}, 
{\em Supergravity Stony Brook workshop, New York, 1979},
ed.\ P. Van Niewenhuizen, D. Freeman (North Holland, Amsterdam, 1980).

\bibitem{Beall:1981ze}
  G.~Beall, M.~Bander, A.~Soni,
  Phys.\ Rev.\ Lett.\  {\bf 48}, 848 (1982).
  


 \bibitem{Maiezza:2010ic}
 For recent complete studies and the references therein, see
  A.~Maiezza, M.~Nemev\v sek, F.~Nesti, G.~Senjanovi\'c,
  Phys.\ Rev.\  {\bf D82 } (2010)  055022:
Y.~Zhang, H.~An, X.~Ji and R.N.~Mohapatra,
  Nucl.\ Phys.\  B {\bf 802}, 247 (2008);


\bibitem{Keung}
  W.-Y.~Keung, G.~Senjanovi\'c,
  Phys.\ Rev.\ Lett.\  {\bf 50 } (1983)  1427.
 

\bibitem{Ferrari} 
  A.~Ferrari {\it et al.},
  Phys.\ Rev.\  D {\bf 62} (2000) 013001;

\bibitem{Gninenko:2006br}
 S.~Gninenko {\em et al.} 
  Phys.\ Atom.\ Nucl.\  {\bf 70} (2007) 441.


\bibitem{KlapdorKleingrothaus:2004wj}
 H.V.~Klapdor-Kleingrothaus {\it et al.},
 Phys.\ Lett.\  {\bf B586 } (2004)  198;
 H.V.~Klapdor-Kleingrothaus, I.V.~Krivosheina,
 Mod.\ Phys.\ Lett.\  {\bf A21 } (2006)  1547.

\bibitem{Mohapatra:1980yp}
  R.N.~Mohapatra, G.~Senjanovi\'c,
  Phys.\ Rev.\  {\bf D23}, 165 (1981).
  

\bibitem{Tello:2010am}
  V.~Tello, M.~Nemev\v sek, F.~Nesti,~G. Senjanovi\'c,~F. Vissani.  
  [arXiv:1011.3522 [hep-ph]], Phys.\ Rev.\ Lett., to appear.


\bibitem{Khachatryan:2010fa} 
  V.~Khachatryan {\it et al.}~[CMS Collaboration]
[arXiv: 1012.5945 [hep-ex]].

\bibitem{Collaboration:2011dx}
  CMS~Collaboration,
  arXiv:1103.0030 [hep-ex].

\bibitem{Abazov:2011xs}
  V.~M.~Abazov {\it et al.} [D0 Collaboration],
  [arXiv:1101.0806 [hep-ex]].




\bibitem{CMSCollaboration:2011wq}
  CMS Collaboration,
  [arXiv:1103.0981 [hep-ex]].


  
\bibitem{Khachatryan:2010mp}
  V.~Khachatryan {\it et al.}~[CMS Collaboration],
  [arXiv: 1012.4031 [hep-ex]] and [arXiv:1012.4033 [hep-ex]].


\bibitem{lq} Using the final states of $\ell\ell jj$ for both
  leptoquark and $W_R$ searches was suggested in
 V.~Bansal,
   arXiv:0910.2215 [hep-ex], and
 M.~Schmaltz, C.~Spethmann,
 arXiv:1011.5918 [hep-ph] who also applied the analysis on Tevatron data.


 
\bibitem{Alwall:2007st}
  J.~Alwall, P.~Demin, S.~de Visscher {\it et al.},
  JHEP {\bf 0709}, 028 (2007).
  [arXiv:0706.2334 [hep-ph]].
   
\bibitem{Sjostrand:2007gs}
  T.~Sjostrand, S.~Mrenna, P.~Z.~Skands,
  Comput.\ Phys.\ Commun.\  {\bf 178 } (2008)  852-867.
  [arXiv:0710.3820 [hep-ph]].  

\bibitem{Hamberg:1990np}
  R.~Hamberg, W.~L.~van Neerven, T.~Matsuura,
  Nucl.\ Phys.\  {\bf B359 } (1991)  343-405.
   
 

 
\bibitem{Cacciari:2005hq}
  M.~Cacciari, G.~P.~Salam,
  Phys.\ Lett.\  {\bf B641 } (2006)  57-61.
  [hep-ph/0512210].

\bibitem{Aaltonen:2010jj}
  T.~Aaltonen {\it et al.} [CDF Collaboration],
   [arXiv: 1012.5145 [hep-ex]].
   
  
\bibitem{Meade:2007js}
  P.~Meade, M.~Reece,
  [hep-ph/0703031].
  
   
\end{thebibliography}
\end{document}